\documentclass{jfm}
\usepackage{graphicx}
\usepackage{natbib}
\usepackage{color}

\def\ADD#1{{{#1}}}    

\def\vv{{\bf v}}

\def\d_M{{\bf d_M}}

\def\rr{{\bf r}}

\def\xx{{\bf x}}
\def\be{\begin{equation}}
\def\ee{\end{equation}}
\def\ba{\begin{eqnarray}}
\def\ea{\end{eqnarray}}
\ifCUPmtlplainloaded \else
  \checkfont{eurm10}
  \iffontfound
    \IfFileExists{upmath.sty}
      {\typeout{^^JFound AMS Euler Roman fonts on the system,
                   using the 'upmath' package.^^J}%
       \usepackage{upmath}}
      {\typeout{^^JFound AMS Euler Roman fonts on the system, but you
                   dont seem to have the}%
       \typeout{'upmath' package installed. JFM.cls can take advantage
                 of these fonts,^^Jif you use 'upmath' package.^^J}%
      }
  \else
  \fi
\fi
\ifCUPmtlplainloaded \else
  \checkfont{msam10}
  \iffontfound
    \IfFileExists{amssymb.sty}
      {\typeout{^^JFound AMS Symbol fonts on the system, using the
                'amssymb' package.^^J}%
       \usepackage{amssymb}%

      }{}
  \fi
\fi

\ifCUPmtlplainloaded \else
  \IfFileExists{amsbsy.sty}
    {\typeout{^^JFound the 'amsbsy' package on the system, using it.^^J}%
     \usepackage{amsbsy}}
    {}
\fi

\newsavebox{\astrutbox}
\sbox{\astrutbox}{\rule[-5pt]{0pt}{20pt}}

\newcommand\p{\ensuremath{\partial}}

\newcommand\eg{e.g.\ }
\newcommand\ie{i.e.\ }

\title[Kolmogorov exact relation for compressible polytropic turbulence]
{A {Kolmogorov-like} exact relation for compressible polytropic turbulence}
\author[Supratik Banerjee and S\'ebastien Galtier]{Supratik Banerjee $^{1,2}$
\thanks{Email address for correspondence: supratik.banerjee@ias.u-psud.fr},\ns and S\'ebastien Galtier $^{1,2}$}

\affiliation{$^1$ Univ. Paris-Sud, Institut d'Astrophysique Spatiale, b\^at. 121, F-91405 Orsay, France \\[\affilskip]
$^2$ Laboratoire de Physique des Plasmas, Ecole Polytechnique, F-91128 Palaiseau Cedex, France}

\pubyear{2010}
\volume{650}
\pagerange{119--126}
\date{\today ; revised ?; accepted ?. - To be entered by editorial office}
\begin{document}

\maketitle

\begin{abstract}
Compressible hydrodynamic turbulence is studied under the assumption of a polytropic closure. Following Kolmogorov, 
we derive an exact relation for some two-point correlation functions in the asymptotic limit of a high Reynolds number. 
The inertial range is characterized by (i) a flux term implying in particular the enthalpy and (ii) a purely compressible term 
${\cal S}$ which may act as a source or a sink for the mean energy transfer rate. At subsonic scales, we predict dimensionally 
that the isotropic $k^{-5/3}$ energy spectrum for the density-weighted velocity field ($\rho^{1/3} \vv$) -- previously obtained for 
isothermal turbulence -- is modified by a polytropic contribution, whereas at supersonic scales ${\cal S}$ may impose another 
scaling depending on the polytropic index. In both cases, it is shown that the fluctuating sound speed is a key ingredient for 
understanding polytropic compressible turbulence. 
\end{abstract}


\section{Introduction}
Turbulence is a ubiquitous phenomenon yet to be understood properly \citep{Frisch95,Sagaut08,Galtier09}. For simplicity 
reasons, incompressible un-magneti\-zed turbulence has received considerable attention and it constitutes naturally the reference 
which other types of turbulence are usually compared to \citep{Kraichnan65,biskamp96,Meyrand12}. The most important 
turbulence property might be summarized by the Kolmogorov's result \citep{Kolmogorov41} which provides an exact relation 
in terms of third-order longitudinal structure functions in the 
asymptotic limit of a very high Reynolds number. This theory is accompanied by a phenomenology which leads 
to a spectral prediction in $k^{-5/3}$ for the energy spectrum. On this basis, one can develop the equivalent of the Kolmogorov's 
exact relation in other contexts, \eg for quasigeostrophic flows or for astrophysical magnetized fluids, or even for a scalar 
passively advected such as a pollutant in the atmosphere 
\citep{Yaglom49, Politano98, Lindborg07, Galtier08, podesta08, Meyrand10,Galtier12}. 

The signatures of turbulence in astrophysical fluids are asserted by {\it in situ} measurements in the solar wind \citep{Carbone09},
or indirect observations like in the solar corona \citep{buchlin} and in the interstellar medium \citep{Elmegreen04}. 
The study of compressible turbulence is fundamental for astrophysical fluid dynamics 
{\citep{Pouquet93,Bhattacharjee98,Marino2010}}. 
For example, it is believed that turbulence acts against the gravitational contraction 
{\citep{Vazquez00}} which, in one galactic cycle, 
may increase the interstellar medium particle density from about $10^{-2}$\,cm$^{-3}$ up to the stellar densities at more than $20$ 
orders of magnitude higher. Additionally, interstellar turbulence can be strongly supersonic like in cold molecular clouds
where turbulent Mach numbers are of the order of $10$ to $100$. Generally speaking, it is thought that compressible turbulence 
controls the rate of star formation and determines the mass distribution of young stars. 

Astrophysical plasmas are often collisionless which questions, in particular, the use of simplistic closure like the barotropic (pressure is a function of the density only) one \citep{belmont}. 
The use of the polytropic approximation is however done for both ions and electrons in the solar wind for the sake of simplicity
\citep{Hu1997, Tu1997}.
In the case of the interstellar medium, for which the collisionality can be weak, a number of studies have nevertheless 
been accomplished to justify that it is polytropic in 
nature within a certain approximation \citep{Hennebelle07, Hennebelle11}. A more realistic approximation is that of 
piecewise polytropic laws where the polytropic index is constant in given intervals of temperatures \citep{Vazquez2003}.

Our understanding of compressible turbulence is significantly narrower than that for the incompressible case. Basic notions on 
the presence of a cascade, an inertial range or  a constant flux energy spectra are still the subject of discussions 
{in the strong turbulence regime} \citep{Aluie11,Aluie13}
{whereas the regime of weak acoustic turbulence -- which is not the subject of the present paper -- was first analysed by 
\cite{Zakharov} who derived an energy spectrum in $k^{-3/2}$ (see also \cite{newell,lvov}).}
In fact our knowledge is mainly limited to direct numerical simulations \citep{Passot87, Lee91, Porter92, Bataille99, Schmidt08,Schmidt09}. 
Results for isothermal turbulence -- using a grid resolution upto $4096^3$ -- \citep{Kritsuk07, Federrath10, Federrath} reveal a 
strong dependence of the energy spectrum on the forcing nature (solenoidal or compressible). A Kolmogorov-like spectrum can 
however be retained for the density-weighted velocity $(\rho^{1/3} \vv)$ if one uses a forcing mainly solenoidal. This result finds 
a natural explanation when it is analysed in terms of the Kolmogorov exact relation derived  in \citet{Galtier11} for compressible 
isothermal turbulence. This exact relation finds another numerical verification in a recent work by \citet{Kritsuk} with direct numerical 
simulations of three-dimensional supersonic turbulence (with a Mach number around $6$); it is also shown that most of the physics for 
supersonic turbulence can be caught with a simplified formulation. 

In the framework of polytropic turbulence, several numerical attempts have been made to date to study the density fluctuations in 
hydrodynamics and compressible magnetohydrodynamics {\citep{Pouquet94,Vazquez96,Vazquez00,Benzi}} along with very 
few works on the energy and velocity spectra. The corresponding theoretical field is considerably lacking thorough analytical works and 
phenomenological views \citep[see \eg][]{Passot88}. Another issue is about the form of the probability distribution function (PDF) 
for the density 
(in practice $\log \rho$) fluctuations. Based on a simple one-dimensional model, it was found that this PDF follows a log-normal law 
in the isothermal case (for which the polytropic index $\gamma=1$) whereas it asymptotically approaches a power law regime at 
high density when $\gamma <1$ and at low density when $\gamma >1$ \citep{Passot98}. The log-normal distribution is symmetric 
(as it is obvious) with respect to its centre (unit fluid density) whereas the power laws are not symmetric about their centers and moreover 
the power laws for $ \gamma > 1$ and $\gamma <1 $ are almost mirror images of one another.
This fundamental difference inside the polytropic regime and between the polytropic and isothermal compressible regimes 
emphasizes the necessity to build a general turbulence theory for polytropic fluids to investigate. 
{Note that for a class of barotropic neutral fluids, where pressure is a function of density only, \cite{Falkovich} have proposed 
a set of generalized exact relations for the current--density correlation function, the Kolmogorov exact relation being a particular limit. This 
general formulation has been tested with direct numerical simulations of isothermal turbulence and surprisingly a lack of universality 
was found \citep{Wagner,Kritsuk}. 
A plausible explanation for this observation is suggested in Appendix \ref{appA}.}

In the present paper, we shall derive an exact relation in terms of two-point correlation functions \ADD{of total energy (which is a 
conserved quantity)} for compressible polytropic turbulence of a neutral fluid. 
\ADD{For the sake of simplicity, the back reaction of the energy equation into the primary variables will be neglected \citep{Passot98}.}
To establish the Kolmogorov-type relation, we shall follow the same formalism as for an isothermal compressible 
fluid \citep{Galtier11} (see also \citet{Banerjee13} for a plasma). The generalization of the previous method to the polytropic case is non 
trivial mainly because: (i) unlike the isothermal case, with a polytropic closure the fluid pressure is no more proportional to the 
fluid density; (ii) the sound speed $C_s$ is no more a constant but varies from point to point in the flow field. Our exact relation 
reveals several new features compared to isothermal turbulence. 
Additionally, the present work has a technical relevance for treating the astrophysical fluids for which the gravitational field plays 
an important role in determining the corresponding dynamics. As we shall explain later, the contribution of the gravitational field to the 
total energy is analogous to the compressible energy part of the polytropic fluid and can be handled in an equivalent manner.

\section{Compressible polytropic fluid}\label{sec:rules_submission}
The basic equations governing the dynamics of a compressible polytropic fluid are:
\ba
\p_t \rho + \nabla \cdot ( \rho \vv ) & = & 0 \, , \label{e1} \\
\p_t (\rho \vv)  + \nabla \cdot (\rho \bold v \otimes \bold v )  &=& - \nabla P  + \bold d + \bold f \, , \label{e2} \\
P &=& K \rho^\gamma \, , \label{e3}
\ea
where $\rho$ is the density, $\vv$ the velocity field, $P$ the pressure, $K$ a constant of proportionality and $\gamma$ the 
polytropic index. The terms $\bold d$ and $\bold f $ represent, respectively, the contributions of the dissipation and the 
external forcing. 
{The latter is assumed to be stationary, homogeneous, delta-correlated in time and acting at large scales only.} 
The sound speed is defined as:
\be
{C_s^2 =  \frac{\partial P}{\partial \rho}                          = {{\gamma P} / \rho}} \, .
\ee

Our analysis will be carried out in the physical space in terms of two-point correlation functions 
and structure functions, where the unprimed quantities represent the properties at the point $\xx$ and the primed quantities 
at the point $\xx'$ (with $\xx' =\xx+\rr$). The fluid is supposed to constitute a statistically homogeneous system which undergoes 
a completely developed turbulence. The analysis is general and does not assume isotropy. Our objective is to set up an exact 
relation associated with the correlators of the total energy density:
\be
E =  {1 \over 2} \rho \vv \cdot \vv  + \rho e \, , 
\ee
where $e$ accounts for the compressible energy which is expressed as:
\be
e = {P \over \rho (\gamma - 1)} = {C_s^2 \over \gamma (\gamma -1)} \, .
\ee

We shall define the two-point correlation functions for the total energy density. The correlator is given by the trace of the matrices 
$ \rho \bold v \otimes \bold v' $ and $ \rho' \bold v' \otimes \bold v $. 
Unlike the isothermal case, here the sound speed is also a flow variable which leads us to write the energy density correlators 
in the following way:
\ba
\left\langle R_E \right\rangle &=& \left\langle  { \rho  } \left( {{\vv \cdot \vv'}\over 2} + 
{{C_s C_s'} \over {\gamma (\gamma -1)}}\right) \right\rangle  , \\
\left\langle R'_E \right\rangle &=& \left\langle  { \rho'  } \left( {{\vv' \cdot \vv}\over 2} +  
{{C_s' C_s} \over {\gamma (\gamma -1)}}\right)\right\rangle   \, .
\ea
Using the above expressions and the statistical homogeneity, one can easily verify that:
\be
\left\langle { {R_E +  R_E'} \over 2 }\right\rangle  = \left\langle{  E  } \right\rangle 
-  \left\langle  {\delta ( \rho \bold v ) \cdot  \delta \bold v \over 4} 
+  {{\delta (\rho C_s) \delta C_s}\over {2\gamma  (\gamma - 1)}} \right\rangle , \label{correl}\\
\ee
where for any variable $\xi$, $\delta \xi \equiv \xi({\bf x} + \rr) - \xi ({\bf x}) \equiv \xi'-\xi$ and $ \langle \cdot \rangle$ 
represents the statistical average.

\section{Derivation of the exact relation}\label{sec:types_paper}
We shall write the partial time derivative of the left-hand side member of Eq. (\ref{correl}). By a straightforward 
calculation, we find:
\ba
{\partial_t} \left\langle  \rho \bold v \cdot \bold v' \right\rangle = &\nabla_\rr& \cdot \left\langle 
- \rho (\bold v \cdot \bold v') \delta \bold v  +  P \bold v'  -  \rho h' \bold v \right\rangle  
+  \left\langle  \rho (\bold v \cdot \bold v' ) (\nabla' \cdot \bold v') \right\rangle + d_1 +  f_1 \, , \label{c1} \\
{\partial_t} \left\langle  \rho' \bold v' \cdot \bold v \right\rangle  = &\nabla_\rr& \cdot \left\langle 
- \rho' (\bold v' \cdot \bold v) \delta \bold v  -  P' \bold v  +  \rho' h \bold v' \right\rangle 
+  \left\langle \rho' (\bold v' \cdot \bold v ) (\nabla \cdot \bold v) \right\rangle + d'_1 + f'_1 \, , \quad  \label{c2}
\ea
where $h$ is the enthalpy ($h=\gamma e$);  {$d_1$, $f_1$ and $d'_1$ and $f'_1$ correspond respectively to the dissipative 
and the forcing terms in Eqs. (\ref{c1}) and (\ref{c2}). Explicit expressions for them can be given as follows:
\begin{eqnarray}
d_1 = \left\langle \textbf{d} \cdot \textbf{v}' + \frac{\rho}{\rho'} \textbf{d}' \cdot \textbf{v}\right\rangle \, , \
f_1 = \left\langle \textbf{f} \cdot \textbf{v}' + \frac{\rho}{\rho'} \textbf{f}' \cdot \textbf{v}\right\rangle \, , \\
d'_1 = \left\langle \textbf{d}' \cdot \textbf{v} + \frac{\rho'}{\rho} \textbf{d} \cdot \textbf{v}' \right\rangle \, , \
f'_1 = \left\langle \textbf{f'} \cdot \textbf{v}+ \frac{\rho'}{\rho} \textbf{f} \cdot \textbf{v'}\right\rangle \, .
\end{eqnarray}
}
We also find:
{
\ba
{\partial_t } \left\langle {{\rho \, C_s C'_s} \over {\gamma (\gamma -1)}} \right\rangle &=& - \left\langle \left( 1 + {{\gamma - 1} \over 2}\right)  {{C'_s C_s}\over {\gamma (\gamma -1)}} 
{\nabla \cdot (\rho \bold v)} \right\rangle 
- \left\langle { \rho \over {2 \rho' \gamma}} C'_s C_s \nabla' \cdot (\rho' \bold v') \right\rangle ,  \, \, \, \, \, \label{c3} \\
{\partial_t } \left\langle {{\rho' C'_s C_s} \over {\gamma (\gamma -1)}} \right\rangle &=& - \left\langle \left( 1 + {{\gamma - 1} \over 2}\right)  {{C_s C'_s}\over {\gamma (\gamma -1)}} 
{\nabla' \cdot (\rho' \bold v')} \right\rangle 
- \left\langle { \rho' \over {2 \rho \gamma}} C_s C'_s \nabla \cdot (\rho \bold v) \right\rangle .  \, \, \, \label{c4}
\ea
}
Adding up Eqs. (\ref{c1}) to (\ref{c4}) and also adding and subtracting the term, 
$\langle (\nabla \cdot \bold v) (\rho' C'_s C_s) + (\nabla' \cdot \bold v') (\rho C_s C'_s)\rangle /\gamma (\gamma -1)$, 
we get by using the definition of the correlators:
\be
\partial_t \left\langle {R_E + R'_E} \right\rangle = 
\ee
$$
\nabla_\rr  \cdot \left\langle - (R_E + R'_E) \delta \bold v 
- {{P' \bold v} \over 2} +  {{P \bold v'} \over 2} + {{\rho' h \bold v'}\over 2} - {{\rho h' \bold v} \over 2} \right\rangle 
+ \left\langle \left( \nabla \cdot \bold v \right)  R'_E + \left( \nabla' \cdot \bold v' \right)  R_E \right\rangle
$$
$$
+ {1 \over {\gamma \left(\gamma -1 \right)}} \left\langle \left(\rho C'_s \vv 
+ \rho' C'_s \bold v \right) \cdot \left( \nabla C_s \right) + \left(\rho' C_s \bold v' 
+ \rho C_s \bold v' \right) \cdot \left( \nabla' C'_s \right) \right\rangle
$$
$$
- {1 \over {2 \gamma}} \left\langle C_s C'_s \left( 1 + {\rho' \over \rho} \right) \nabla \cdot \left(\rho \bold v \right) 
+ C_s C'_s  \left( 1 + {\rho \over \rho'} \right) \nabla' \cdot \left(\rho' \bold v' \right) \right\rangle + \cal D + \cal F \, , 
$$
\ADD{where ${\cal D} = (d_1 + d'_1)/2$ and ${\cal F} =  (f_1 + f'_1)/2$ represent, respectively}, the resultant dissipative and forcing terms. 
By introducing in the above expression relation (\ref{correl}) without the statistical average, we obtain eventually:
\be
\partial_t \left\langle {R_E + R'_E} \right\rangle = 
\ee
$$
\nabla_\rr
\cdot \left\langle {1 \over 2} \left( \delta \left(\rho \bold v \right) \cdot \delta \bold v \right) \delta \bold v 
+ {1 \over {\gamma (\gamma-1)}} \delta \left(\rho C_s \right)  \delta C_s \delta \bold v \right\rangle 
+ \cal D + \cal F 
$$
$$
+ \left\langle \left( \nabla \cdot \bold v \right) \left( R'_E - E' + {P' \over 2} \right) + \left( \nabla' \cdot \bold v' \right) 
\left( R_E - E + {P \over 2} \right) \right\rangle
$$
$$
+ {1 \over {\gamma \left(\gamma -1 \right)}} \left\langle \left(\rho C'_s \bold v 
+ \rho' C'_S \bold v \right) \cdot \left( \nabla C_s \right) + \left(\rho' C_s \bold v' 
+ \rho C_S \bold v' \right) \cdot \left( \nabla' C'_s \right) \right\rangle
$$
$$
+  \left\langle \left[ {{C'_s}^2 \over {2 \left(\gamma - 1\right)}} - {{C_s C'_s} \over {2 \gamma}} 
\left( 1 + {\rho' \over \rho} \right) \right]  \nabla \cdot \left(\rho \bold v \right)  
+ \left[ {{C_s}^2 \over {2 \left(\gamma - 1\right)}} - {{C_s C'_s} \over {2 \gamma}} 
\left( 1 + {\rho \over \rho'} \right) \right] \nabla' \cdot \left(\rho' \bold v' \right) \right\rangle . 
$$
Now we introduce the usual assumptions specific to three-dimensional fully developed turbulence with a direct energy cascade 
\citep{Frisch95}. We consider a steady state for which the partial time derivative of the average energy correlators vanishes. 
\ADD{
We consider a small enough viscosity such that the dissipative term will not affect the inertial range. For incompressible turbulence 
the dissipation is a sink localized mainly at the smallest scales of the system but in the present situation this property is not 
guaranteed. 
For example with the one dimensional Burgers equation -- a simple archetype equation for very high Mach number flows -- the 
contribution of the dissipation term is not concentrated at small scales but is rather constant throughout the whole inertial range. 
Its value tends to zero only as the viscosity goes to zero. Note that this is true for regular shocks but might even become wrong for 
shocks of Alfvenic type where dissipation may affect large scales as it shown in one dimensional simulations \citep{Laveder}.}
The mean energy injection rate is determined by the resultant forcing which is in fact, under our assumptions, 
\ADD{${\cal F} =  2 \varepsilon$} \citep{Galtier11}. 
\ADD{Note that the question of a forcing acting at large scales only has been discussed recently in \cite{Kritsuk} in a 
numerical context for which it is not an obvious implementation.}
Then, far in the inertial zone (infinite Reynolds number limit is assumed) where the dissipative terms are negligible \citep{Aluie13}, the 
exact relation writes:
\be
- 2 \varepsilon = \nabla_\rr \cdot \left\langle {1 \over 2} \left( \delta \left(\rho \bold v \right) \cdot \delta \bold v \right) \delta \bold v 
+ {1 \over {\gamma (\gamma-1)}} \delta \left(\rho C_s \right)  \delta C_s \delta \bold v
 + {{ \overline{\delta} h} \delta ({\rho \bold v}}) \right\rangle \label{er}
 \ee
$$
+ \left\langle D \left( R'_E - E'+ {P' \over 2}- {1 \over \gamma} \overline{\delta} \rho \, C_s C'_s \right) 
+ D'  \left( R_E - E + {P \over 2}-  {1 \over \gamma} \overline{\delta} \rho \, C_s C'_s\right) \right\rangle \, , 
$$
where $\overline{\delta} \xi \equiv (\xi+\xi')/2$ and $D$ and $D'$ denote respectively $ (\nabla \cdot \vv) $ and $ (\nabla' \cdot \vv ' ) $. 
Note that in the derivation we have used the relation, 
$\left( \bold v \cdot \nabla \right) C_s = ({{(\gamma-1)C_s} / {2 \rho}}) \bold v \cdot \nabla \rho$.

Expression (\ref{er}) is our main result: it is an exact relation for three dimensional compressible polytropic turbulence. 
It is composed of the divergence of a flux ${\bf F}$ (first line in the right hand side) and of a purely compressible term 
${\cal S}$ (second line) which leads us to use for the discussion the simplified writing:
\be
-2 \varepsilon = \nabla_\rr \cdot {\bf F} + {\cal S}(\rr) \, . 
\label{compact}
\ee
\ADD{As for isothermal turbulence, ${\cal S}$ may be seen as a source or a sink for the mean energy transfer rate. But unlike the 
isothermal case, here the determination of the sign of the source term is not immediate in general and depends on the competition 
between $R'_E -E'$ (which is mainly negative) and $\overline{\delta} \rho C_s C'_s/ \gamma - {P'}/{2}$
(whose sign is more difficult to define although it is positive at small scales since it tends to $P/2$ when $r \to 0$). 
Thus, ${\cal S}(r)$ contributes to modify $\varepsilon$ for giving an effective mean total energy injection rate 
$\varepsilon_{\rm{eff}}$ (with $\varepsilon_{\rm{eff}} \equiv \varepsilon+ {\cal S}/2$) possibly larger than $\varepsilon$ in the 
compression case and smaller than $\varepsilon$  in the dilatation case with possibly an inverse cascade if 
$\varepsilon_{\rm{eff}} <0$. 
An illustration of dilatation and compression effects in the space correlation is given in \cite{Galtier11} (see Fig. 1). }

\section{Discussion}\label{sec:filetypes}
\subsection{Incompressible limit}

First of all, let us check the incompressible limit of a polytropic fluid, \ie $ \gamma \rightarrow + \infty $. For that limit, we obtain 
$D = 0 $ and a uniform density at every point of the flow field. We also get for the second term of the flux, 
$\delta (\rho C_s) \delta C_s  \sim C_s^2  \sim \gamma $ (as all of them tend to infinite value), which does not lead to a singularity 
thanks to the presence of ${\gamma (\gamma -1)}$ in the denominator. 
\ADD{The third term also goes away in the incompressible limit under the following justification:
\be
\nabla_\textbf{r} \cdot \left\langle \overline{\delta} h \delta (\rho \textbf{v}) \right\rangle  
= \nabla_\textbf{r} \cdot \left\langle \gamma \overline{\delta} e \delta (\rho \textbf{v}) \right\rangle 
= \nabla_\textbf{r} \cdot \left\langle \frac{\gamma}{2(\gamma -1)} \left( \frac{P}{\rho} + \frac{P'}{\rho '}\right) \delta (\rho \textbf{v}) \right\rangle \, .
\ee
In the limit where $\gamma \to +\infty$, we have $\rho = \rho' = {\rm{constant}}$ and we get (using $ D= D'= 0 $):
\ba
\nabla_\textbf{r} \cdot \left\langle \overline{\delta} h \delta (\rho \textbf{v}) \right\rangle  
&=&  \frac{1}{2} \nabla_\textbf{r} \cdot \left\langle  P' \vv ' + P \vv ' - P' \vv - P \vv \right\rangle \nonumber \\
&=& \frac{1}{2} \left\langle P \left( \nabla' \cdot \vv ' \right)  + P' \left( \nabla \cdot \vv  \right) \right\rangle   = 0 \, .
\ea}
The term ${\cal S}$ vanishes automatically due to the solenoidal velocity 
field and the uniform density. Then, the Kolmogorov's exact relation is reproduced properly \citep{antonia97}. 
\ADD{Note that our exact relation (\ref{er}) for compressible turbulence implies third-order correlators like in the incompressible
case.}

\subsection{Dimensional analysis and spectra}

\ADD{To start with the spectral prediction, we keep the source term ${\cal S}$ aside and investigate what happens dimensionally 
for the spectral prediction just with the flux term (isotropy is assumed). Additionally, we will not consider any intermittency correction 
which can modify slightly our conclusion about the scaling laws. At this point of discussion, it is necessary to justify the scale 
invariance of the mean total energy density injection rate $ \varepsilon$ in the compressible case. According to \citet{Falkovich} 
even in compressible turbulence a scale invariant mean energy flux rate can be assumed if the forcing correlation length scale is 
much larger than the inertial range length scales. A discussion around this question has been developed in \citet{Wagner} and 
\cite{Kritsuk} where it is claimed that a very short time correlation for the large scale acceleration or a very small length correlation  
for the density functions is necessary for the scale invariance of $\varepsilon$. 
Under this assumption, the exact relation can be written mainly as:
\be
- 2 \varepsilon \simeq {{ \left( \rho v\right) _\ell v_\ell^2}\over \ell} \left( \frac{1}{2} + {1\over{ {\gamma(\gamma -1)}  {M_\rho}_\ell M_\ell} } + \frac{1}{\left( \gamma - 1\right) {\cal M}_\ell^2} +  \frac{1}{ 4 \left( \gamma - 1\right) {M}_\ell ^2}        \right) \, , \label{s1}
\ee
where:
\be
{M_\rho}_\ell \equiv { {\delta \left( \rho v \right) } \over {\delta \left( \rho C_s \right) } }  \sim { \left( \rho v \right) _\ell \over {\left( \rho C_s \right) }_\ell} , 
\quad
M_\ell \equiv { {\delta v} \over {\delta C_s} } \sim {v_\ell \over {C_s}_\ell} , 
\quad
{\cal M}_\ell \equiv {\delta v \over {\overline {\delta} C_s}}  \sim {v_\ell \over  {C_s}} ,  
 \ee
are respectively the {\it current} Mach number, the  {\it gradient} Mach number (which is not defined for isothermal 
turbulence where the sound speed is constant) and the {\it turbulent} Mach number. The third one is familiar in turbulence studies 
whereas the first and the second one have been defined for the sake of our current study. It is not obvious to built up any spectral 
assumption from the above expression (\ref{s1}). Insofar as we assume further simplications like 
$(\rho v)_\ell v_\ell^2 \sim \rho_\ell v_\ell^3 $ and 
$ { \left( \rho v \right) _\ell / {\left( \rho C_s \right) }_\ell} \sim {v_\ell / {C_s}_\ell} $, we can approximately write:
\be
-  4 \varepsilon \simeq {{ \rho_\ell  v_\ell^3}\over \ell} \left( 1 + \frac{\left( \gamma + 4\right)} {{ {2\gamma(\gamma -1)}   M_\ell^2} } 
+ \frac{2}{  \left( \gamma - 1\right) {\cal M}_\ell^2}  \right) \, . \label{s2}
\ee 
Additionally, if we assume that $M_\ell \sim \ell^\alpha$ and $\cal M_\ell \sim \ell^\beta$, expression (\ref{s2}) can be re-written as:
\be
 - 4 \varepsilon \sim {{ \rho_\ell  v_\ell^3}\over \ell} \left( 1 + \Gamma_1 \ell^{- 2 \alpha} + + \Gamma_2 \ell^{- 2 \beta} \right) \, ,
\ee
with the coefficients $ \Gamma_1 = {\left( \gamma + 4\right)} / [{2\gamma(\gamma -1)}] $ and $ \Gamma_2 = {2}/{(\gamma-1)}$. 
One can easily verify that $\Gamma_1 \sim \Gamma_2$ and so  none of the second and the third terms can be neglected with 
respect to one another just from their coefficient consideration. From this step after some straightforward calculations, one can 
predict that the power spectrum of density-weighted velocity $ w \equiv \rho^{{1}/{3}} v $ scales as:
\be
E_k^w \sim \varepsilon^{2\over 3} k^{-{ 5 \over 3}} \left( 1 + \Gamma_1 k^{2 \alpha} + \Gamma_2 k^{2 \beta} \right) ^{-{2\over 3}} \, . \label{pred1}
\ee}

\ADD{For supersonic turbulence for which $ \delta v \gg \overline{\delta} C_s $ and $ \delta v \gg \delta C_s $, the second and the third 
terms become negligible compared with the first one and we have:
\be
E_k^w \sim {\varepsilon_{\rm{eff}}}^{2/3} k^{-{5 / 3}} , 
\ee 
whereas for $ \delta v \gg \overline{\delta} C_s $ but $ \delta v \ll \delta C_s $ (which is less probable but still possible), we have:
\be
E_k^w \sim \left( {\varepsilon_{\rm{eff}} / \Gamma_1}\right) ^{2/3} k^{-{\left( 5 + 4 \alpha \right) \over 3}}  ,
\ee 
where $\varepsilon_{\rm{eff}}$ reflects the non-negligible effect of the source terms in the supersonic turbulence regime (see next 
subsection).}

\ADD{For subsonic turbulence we may have two possible situations. First, we have the case $ \delta v \ll \overline{\delta} C_s$ but 
$ \delta v \gg \delta C_s $ for which the spectral relation takes the form:
\be
E_k^w \sim \left( {\varepsilon / \Gamma_2}\right) ^{2/3} k^{-{\left( 5 + 4 \beta\right) \over 3}}  .
\ee 
One may immediately notice that if the scale dependence of the gradient or the turbulent Mach number is weak, \ie $\alpha$ or 
$\beta$ takes a small value, the power spectrum for $w$ tends to the Kolmogorov value. Finally, when $ \delta v \ll \overline{\delta} C_s $ 
and $ \delta v \ll \delta C_s $, we are left with:
\be
E_k^w \sim \varepsilon^{2\over 3} k^{-{ 5 \over 3}} \left( \Gamma_1 k^{2 \alpha} + \Gamma_2 k^{2 \beta} \right) ^{-{2\over 3}} \, , \label{pred1}
\ee
which is no more a pure power law but a non-trivial combination of two power laws. A power law can nonetheless be recovered if 
$\alpha \simeq \beta$. Note that the above analysis cannot be used in the isothermal limit as $\Gamma_1$ and $\Gamma_2$ are undefined for 
$\gamma = 1$. The basic reason for this problem is our total energy expression whose compressive part is undefined in the isothermal case and 
cannot be obtained as a limit of a polytropic case (for which $\gamma \to 1$).}

\subsection{Source term contributions}

The contribution of ${\cal S}$ is expected to be non negligible at supersonic (${\cal M}_\ell \gg 1$) scales. 
An intuitive argument for this can be found in \cite {Biskamp} where the dilatation term $D$ is shown to be approximately 
proportional to the Mach number squared. Following the same formalism as carried out by \cite{Kritsuk} for isothermal turbulence, 
we may rewrite the polytropic source term as:
\be
{\cal S} = \left \langle \frac{1}{2} \left[ \delta (\rho \vv D) - 2 \overline{\delta} D \delta (\rho \vv) \right] \cdot \delta \vv 
+ \left[ \frac{\delta(\rho C_s D)}{\gamma(\gamma - 1 )} 
+ \frac{(\gamma - 3)}{\gamma(\gamma-1)} \overline{\delta} D \delta (\rho C_s) \right] \delta C_s - PD \right \rangle \, . \label{source}
\ee
In the subsonic case, irrespective of the sub-regimes, $ C_s ( \sim \overline{\delta} C_s) $ is larger than 
$ \delta v $ and $ \delta C_s $ and 
hence the source term comes to be simply $ \langle - PD \rangle $. This expression, bereft of any fluctuation, can hardly be expected to 
participate in turbulence and spectral construction, which in turn justifies why for subsonic turbulence the basic contribution is from flux terms.  

On the other hand, for supersonic regime (with moderate $\gamma$) where  $\delta v \gg C_s $ and $ \delta v \gg \delta C_s$ , 
the source term is reduced to:
\be
{\cal S} = \left \langle \frac{1}{2} \left[ \delta (\rho \vv D) - 2 \overline{\delta} D \delta (\rho \vv) \right] \cdot \delta \vv  \right \rangle \, .
\ee
\ADD{This expression is similar to Eq. (2.10) in \cite{Kritsuk} where a relatively small contribution has been found numerically for 
isothermal supersonic turbulence. We can therefore conclude that in the supersonic turbulence regime $({\cal{M}}_\ell >1)$ the source  
is weakly affected by the polytropic terms at moderate values of $\gamma$. This point can be quantified numerically for polytropic turbulence 
by comparing the relative importance of each term in ${\cal S}$.}

\subsection{Different values of polytropic index $\gamma$}

For the discussion, we shall consider the isothermal case \citep{Galtier11} as a reference against which the polytropic law will 
be compared. We see that the polytropic closure leads to the appearance of new terms in the flux and the source.
\ADD{From expression (\ref{s2}), one can immediately see that the contribution of the second and third terms in the flux may enhance that of 
the first one for $\gamma >1$ but opposes when $\gamma <1$. More precisely, for $\gamma <1$ we may expect even the possibility of 
an inverse cascade of total energy if the first term becomes subdominant which practically corresponds to Eq. (\ref{pred1}). 
From a theoretical point of view this situation may arise at low gradient and turbulent Mach number for which the sound speed and its 
fluctuations are relatively large with respect to the velocity fluctuations of the fluid. This property could be investigated numerically by looking 
at the relative importance of each term inside the flux. Besides $\gamma =1$ (which is discussed above) the flux contains another singularity 
for $\gamma=0$ due to the presence of the second term in Eq. (\ref{s2}).}

For the source terms the effect of $\gamma$ is subtle. The second term of expression (\ref{source}) -- the one multiplied by 
$\delta C_s$ -- depends on the $\gamma$ values. For $\gamma >3$, both members of the second term have positive coefficients. 
For $1 < \gamma < 3$, the coefficient of the first member ($1/[\gamma(\gamma-1)]$) is positive whereas it is negative for the second 
one ($(\gamma-3)/[\gamma(\gamma-1)]$). 
If $ 0< \gamma <1$, the opposite case to the previous one will occur. For astrophysical interest, it is however possible to get 
negative values of 
gamma too \citep{Horedtbook}. For that situation, the first and second terms may contribute with a different sign. 
In order to verify numerically these effects, it is needed to consider a flow with very low gradient Mach number $(M_\ell)$ for 
which the first term (multiplied by $\delta v$) of the source contributes weakly with respect to the second term of the source (multiplied by 
$\delta C_s$). At the same time, it is also essential to weaken the effect of the third term \ie $\left\langle -PD \right\rangle$ which is probably 
not obvious to satisfy. 
In reality, the case $\gamma<0$ corresponds, in general, to the thermal instability in the outer envelopes of giant molecular clouds \citep{renard93} and therefore requires a more complicated model.

\section{Conclusion}
In the present paper, a Kolmogorov exact relation is derived for three-dimensional polytropic hydrodynamic turbulence. 
This result generalizes the isothermal relation obtained recently \citep{Galtier11} and emphasizes the importance of polytropic effects 
in compressible turbulence. In particular, an asymmetry is found in the exact relation in comparison with the value of the polytropic index. 
This observation is somewhat analogous (by nature) to the results obtained by \citet{Passot98} with a one-dimensional model where a 
remarkable difference was found in the PDF of $\log \rho$ which asymptotically approaches (getting away from $ \gamma =1 $)  an 
asymmetric power law regime at high density when $\gamma <1$ and at low density when $\gamma >1$. Further investigations are 
clearly needed -- probably with direct numerical simulations -- to better understand this regime. 

The present work shows that the gradient Mach number and its scaling play a fundamental role in the determination of 
the density-weighted velocity spectrum in subsonic polytropic turbulence with a moderate polytropic index. Depending on that, 
we define two sub-regimes in subsonic turbulence. The construction of a density power spectrum is feasible from both the 
astrophysical observations \citep[\eg][]{Armstrong81} and numerical simulations \citep{Kim05} whereas its 
construction from a statistical exact relation is non trivial. The reason can readily be understood if we explain it in terms 
of pressure power spectrum. The compressible fluid pressure does not obey a Laplace equation unlike the incompressible case. 
Thus, the pressure scaling cannot be related to the velocity power spectrum or even to compressible energy spectrum, which 
finally prevents us from obtaining a scaling relation for density fluctuations despite having the closure relations. The attempt to 
construct the density fluctuation spectrum directly from the mass conservation has also been examined and has not seemed to 
give expected result.

The current formalism can be extended to polytropic fluids under a gravitational field which has some interest in astrophysics. 
In this situation, the continuity equation does not change, the Navier-Stokes equations consist in a supplementary term due to 
the gravitational intensity and the Poisson's equation relates the gravitational potential to the fluid density. 
We think that the technics used in the present derivation can be adapted directly to this problem since the polytropic and gravitational 
contributions appear to have the same nature. The effect of gravity is, however, non trivial because of the Poisson's equation and so is 
the negative contribution of the gravitational term to the total energy. 
A detailed work on this aspect is in preparation and will be presented elsewhere.


\section*{Acknowledgement}
We acknowledge A. Kritsuk, R. Grappin and P. Hennebelle for useful discussions and suggestions. We would like to give a special thanks to C. Federrath for his useful comments for improving our article. Finally, a sincere gratitude to all the three referees who have carefully read and given constructive comments and suggestions for further refinement of our article.

\appendix
\section{}\label{appA}
\ADD{
This Appendix is devoted to the comparison between our derivation and the one proposed by \citet{Falkovich} where an exact 
relation for the current-density correlation function was proposed. From Eqs. (\ref{e1})--(\ref{e2}) we obtain:
\ba
{\partial_t} \left \langle  \rho \rho' \bold v \cdot \bold v' \right\rangle &=& 
\left \langle \rho \bold v \cdot {\partial_t} (\rho' \bold v') \right\rangle + 
\left \langle  \rho' \bold v' \cdot {\partial_t} (\rho \bold v) \right\rangle \nonumber \\
&=& - \nabla_\rr \cdot \left\langle 
\rho \rho' (\bold v \cdot \bold v') \delta \bold v  +  \rho P' \bold v  -  \rho' P \bold v' \right\rangle 
+ \tilde d +  \tilde f \, , \label{falk1}
\ea
where:
\be
\tilde d = \tilde d(\rr) = \left\langle \rho' \textbf{v}' \cdot \textbf{d} + \rho \textbf{v} \cdot \textbf{d}' \right\rangle  \, , \quad
\tilde f = \tilde f (\rr) = \left\langle \rho' \textbf{v}' \cdot \textbf{f} + \rho \textbf{v} \cdot \textbf{f}' \right\rangle \, .
\ee
We have the following relationship for homogeneous turbulence: 
\be
\left\langle \delta (\rho \bold v) \cdot \delta (\rho \bold v) \right\rangle = 
2 \left\langle \rho^2 \bold v^2 \right\rangle - 2 \left\langle \rho \rho' \bold v \cdot \bold v' \right\rangle \, ,
\ee
which leads to:
\be
\partial_t \left\langle \rho \rho' \bold v \cdot \bold v' \right\rangle = 
{1 \over 2} \partial_t \left\langle \delta (\rho \bold v) \cdot \delta (\rho \bold v) \right\rangle
- \partial_t \left\langle \rho^2 \bold v^2 \right\rangle \, . \label{app1}
\ee
Since the quantity $\rho^2 \bold v^2$ is {\it not} an inviscid invariant its time derivative introduces a nonlinear contribution 
which has a non conservative form, namely:
\be
\partial_t \left\langle {\rho^2 \bold v^2} \right\rangle = 
- 2 \left\langle \rho \bold v \cdot [\nabla \cdot (\rho \bold v \otimes \bold v)] + \rho \bold v \cdot \nabla P \right\rangle
+ \tilde d(0) + \tilde f(0) \, . 
\ee
In the derivation made by \citet{Galtier11} the use of an inviscid invariant -- the total energy -- does not lead to the appearance of such type 
of nonlinear contribution. Therefore, it is important to check (\eg by numerical simulations) if the assumption of stationarity can be applied 
to the current-density correlation function (whereas it is applicable for the fluctuating part written in terms of structure functions). 
Otherwise, expression (\ref{falk1}) becomes in the inertial range: 
\ba
\partial_t \left\langle {\rho^2 \bold v^2} \right\rangle + \bar \varepsilon &=& \nabla_\rr \cdot \left\langle 
\rho \rho' (\bold v \cdot \bold v') \delta \bold v  +  \rho P' \bold v  -  \rho' P \bold v' \right\rangle  \nonumber \\
&=& \nabla_\rr \cdot \left\langle 
\rho \rho' (\bold v \cdot \bold v') \delta \bold v  - 2 \delta (\rho \bold v) \bar \delta P \right\rangle \, , \label{falk2}
\ea
where $\bar \varepsilon = \tilde f(0)$ is the injection rate of momentum squared (see \citet{Falkovich}). 
In the context of isothermal turbulence, direct numerical simulations \citep{Wagner,Kritsuk} have shown that the classical derivation 
based on the inviscid invariant gives better results with the possibility to detect universality. It is believed that the previous arguments
could be the explanation for the lack of universality in the \citet{Falkovich} paper. 
}

\bibliographystyle{jfm}
\bibliography{Supranew}
\end{document}